\documentstyle[prl,aps,epsf,epsfig,multicol]{revtex}    % Specifies the document style.

\begin{document}
%\include{psfig}
%\parskip=10.pt
%\baselineskip=20.pt
\bibliographystyle{plain}
\title{Orientational Ordering in Sequence-Disordered Liquid Crystalline Polymers}
\author{Lorin Gutman and Eugene Shakhnovich}
\address{Department of Chemistry and
Chemical Biology, Harvard University, Cambridge, MA 02138}
\maketitle
\begin{abstract}
Phase separation of sequence-disordered liquid crystalline polymers, 
a promising class of technological and biological relevance,
is studied by field theory, and thermodynamic mechanisms responsible
for orientational ordering observed in experiments, are discussed.
The theory developed predicts that chemical disorder marginally affects
the nematic/isotropic biphasic coexistence width, but strongly impacts
ordering; above a critical chemical disorder threshold orientational
ordering is precluded.
%\noindent ({\it submitted to PRL})
\end{abstract}
%\baselineskip=20.pt
%\parskip=10.pt
%\include{psfig}
\vspace{-.8cm}
\begin{multicols}{2}
The study of sequence heterogeneity effects on phase separation of
sequence-disordered liquid crystalline polymers is paramount to
modulation of electro-optical properties, piezo and pyro-electricity
\cite{kn:gnoptical}, and also to disclosure of general ordering trends
prevalent in proteins with secondary structure \cite{kn:gnproteins}.
Onsager approach \cite{kn:gnOnsager}, developed for the study of liquid
crystals (LC's) with repulsive anisotropic interactions, has been
extended and used \cite{kn:gnGrosberg} to analyze single chains and
melts of liquid crystalline homopolymers (LCP's).  The effect of
attractive anisotropic interactions by dipolar coupling was first
introduced in the theory of LC's by Mayer and Saupe \cite{kn:gnMayer};
their approach was found useful in the study of low molecular weight
LC's.  LCP's were also studied by lattice models \cite{kn:gnFlory},
and the Onsager approach was used to correct for lattice artifacts
effects in phase behavior \cite{kn:gnRonca}.

A field theoretic approach to study many chain LCP's was proposed by
Gupta and Edwards (GE) \cite{kn:gnEdwards}; free energy contributions
to nematic ordering were computed non-perturbatively, and the Landau
expansion method was avoided by summation over all Rouse modes;
interaction-wise, both athermal repulsive contributions
\cite{kn:gnOnsager}, and the Mayer Saupe dipole-dipole attractive
contributions to the inter-segment interaction potentials are
present. GE \cite{kn:gnEdwards} predict that the nematic/isotropic
(N/I) transition is first order while the transition temperature is
depressed in comparison with low molecular weight rods; their
predictions are supported by computer simulations of short nematic
chains \cite{kn:gnFrenkel} and by experiments
\cite{kn:gnAllest}. Recently, GE approach was employed to study
orientational ordering in semi-flexible homopolymers embedded in
flexible surfaces \cite{kn:gnPodgornich}.

Stupp et al. \cite{kn:gnoptical}, \cite{kn:gnStupp} synthesized and
characterized by optical microscopy and C13 NMR, orientation and
sequence-statistics of thermo-tropic sequence-disordered LCP. In these
experiments, optical domains were observed over a broad temperature
regime $\sim 120^{o}$; these domains were attributed to a wide
biphasic N/I coexistence width, and their occurrence was attributed to
sequence-disorder and finite chain length effects.  Fredrickson et
al. \cite{kn:gnFred}, studied via Landau theory the homogeneous chain
anisotropy limit of the sequence model \cite{kn:gnGutin}, and
predicted that an increase in chain length diminishes chemical
heterogeneity effects on the N/I biphasic width.

In order to make closer connection with experiments, in the present
work we construct a field theory of LCPs made of stiff mesogens and
flexible spacers, randomly distributed on the chain (viz. fig. 1).
Sequence-distribution and interaction-wise, our model is an adequate
description of the sequence disordered LCPs synthesized in experiments
\cite{kn:gnStupp2}.  While the GE method is not applicable here, in
the homopolymer limit the present study reproduces the free energy of
GE theory \cite{kn:gnEdwards}.

The many-chain Hamiltonian for a solution of sequence disordered LCP's
made of mesogenic segments (A's) and flexible segments (B's) is:  
\begin{eqnarray}
H=\nonumber\\ \sum_{i} \int (\frac{1-\theta(n_{i})}{2}\frac{3}{2l}{\bf
u}^{2}(n_{i}) + \frac{1+\theta (n_{i})}{2}\frac{\beta \epsilon}{2kT}
\dot{\bf u}^{2}(n_{i})) dn_{i}\nonumber\\ -w \int d{\bf r}
[\hat{\sigma}^{ii}({\bf r}) \hat{\sigma}^{jj}({\bf r})-
\hat{\sigma}^{ij}({\bf r})\hat{\sigma}^{ji}({\bf r})] -u \int d {\bf
r} \hat{\rho}({\bf r})^{2} \label{eq:sthamiltonian}
\end{eqnarray}

\begin{eqnarray}
\hat{\rho({\bf r})}=\sum_{k}\int dn_{k}\delta( {\bf r}-{\bf r}(n_{k}))\nonumber\\ 
\hat{\sigma}^{ij}({\bf r})=\sum_{k}\int dn_{k}\frac{1+
\theta(n_{k})}{2} \delta({\bf r}-{\bf r}(n_{k})) {\bf
u}_{k}^{j}(n_{k}){\bf u}_{k}^{i}(n_{k}) \label{eq:lagrange}
\end{eqnarray}
${\bf r}(n_{i})$ is the spatial location of the $ n $'th segment of the
i'th chain, ${\bf u}(n_{i})$ is the chain tangent of the i'th chain at
$n_{i}$ and $\theta(n_{i})$ is the chemical composition variable of
the $n_{i}$'th segment on the i'th chain.  $\theta(n_{i})=1$ for an A
segment and $\theta(n_{i})=-1$ for a B segment.  The sequence
heterogeneity, represented by fluctuations in the segment composition
along the chain contour obeys a Gaussian process with mean,
$<\theta>=2f-1$, $\delta \theta= \theta(n_{i})-<\theta>$, and sequence
fluctuations
$\overline{\delta \theta^{2}}=<(\delta\theta(n_{i})\delta\theta(n_{j}'))>$=$\delta(n_{i}-n_{j}')4f(1-f)l$
\cite{kn:gngutmanarup}; f is the fraction of A segments, and l is the
statistical segment length.  The first term in
eq. \ref{eq:sthamiltonian} is the nearest-neighbors harmonic
interaction potential of flexible segments of type B, while the second
term in eq. \ref{eq:sthamiltonian} precludes bending of the
stiff segments, A's, (viz. line-bounded mesogens, fig. 1).  
The third term in eq. \ref{eq:sthamiltonian} is the
interaction potential of A-A segments by anisotropic (viz. also
\cite{kn:gnEdwards}) alignment of A pairs; this energetic penalty is
zero for aligned tangents of A segments adjacent in space, and w for
normal tangents. 
\begin{figure}[ht]
\vspace{-2.cm}
\psfig{file=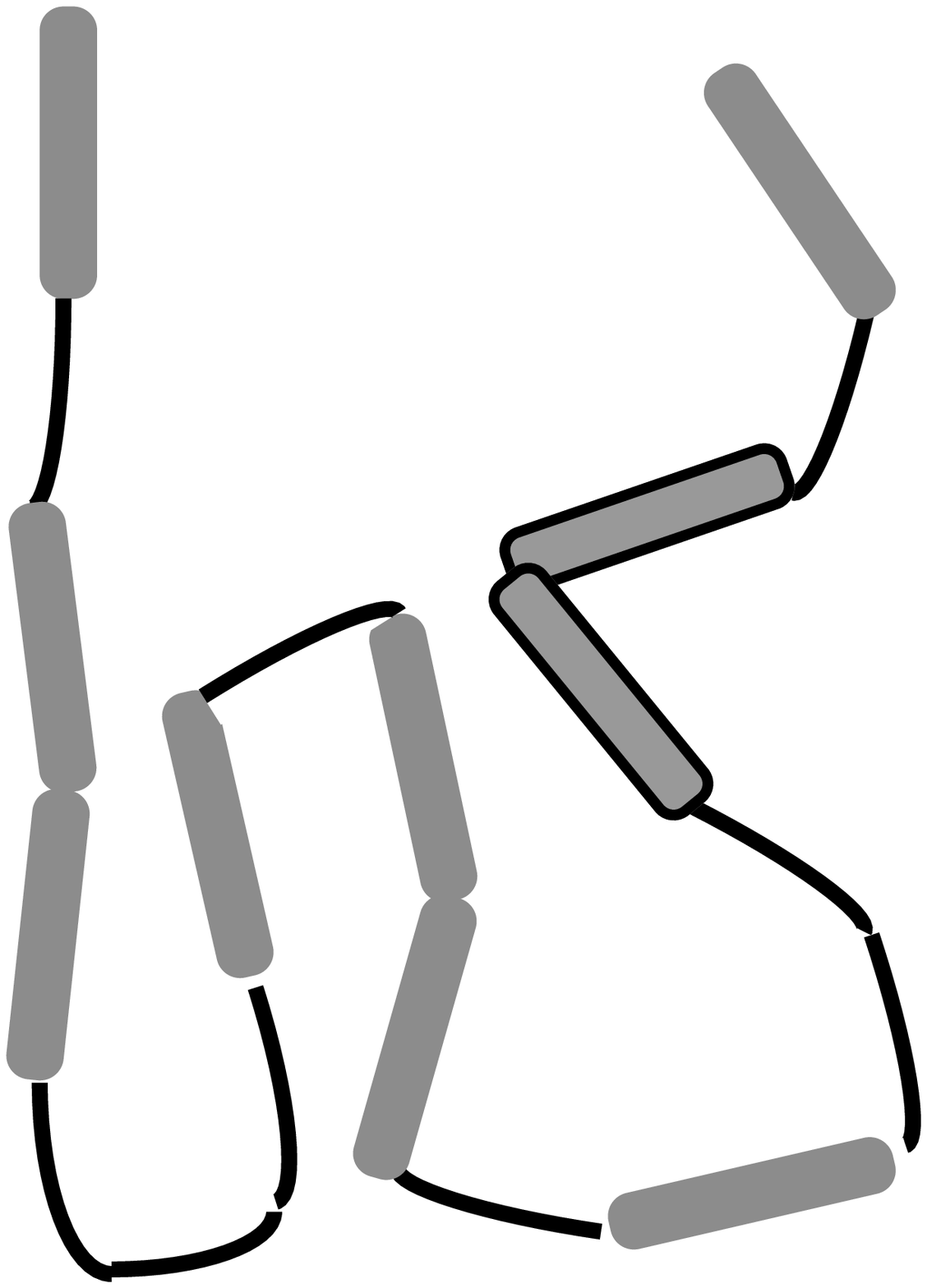,width=6.cm}
\vglue-0.2cm
\vspace{0.4cm}
{\small Fig. 1: 
One disordered LCP sequence. Flexible segments are thin and black, mesogen segments
are thick and grey}
\label{cv72b}
\end{figure}
\vglue.0cm
\vspace{-0.2cm}
The A-A anisotropic interaction potential, w,
contains both athermal and thermo-tropic contributions
\cite{kn:gnEdwards}. The fourth term in eq. \ref{eq:sthamiltonian}, is
the excluded volume inter-segment interaction. While sequence effects
on equi-stiff heteropolymers were treated \cite{kn:gnFred}, the
present work focuses on the effect of dissimilarity in anisotropy
among the stiff mesogens (A) and the flexible spacer (B) on
orientational ordering, in lyotropic and thermo-tropic
sequence-disordered LCPs.

Below, solution steps are briefly described. First, continuous microscopic
orientational tensors are introduced by delta function
constraints on the partition function, while the fluctuations of the chain
director of the stiff segments is constrained in the usual way (viz. \cite{kn:gnEdwards}
for further details). The partition function at fixed sequence is given by:
\begin{eqnarray}
Z[\theta(n_{i}),{\bf u}(n_{i})]=\underline{\int}\overline{\int}[\prod_{n_{i},{\bf r}} D{\bf
r}_{i}(n_{i}) D \sigma({\bf r})] exp(-L[\hat{\sigma}({\bf r})])\nonumber\\ 
\prod_{n_{i},{\bf r}}\delta[\sigma({\bf r})-\hat{\sigma}({\bf r})
\delta[{\bf u}(n_{i})^{2}\frac{(1+\theta(n_{i}))}{2}-1] 
\label{eq:partition}
\end{eqnarray}
$\hat{\sigma}({\bf r})$ is given in eq. \ref{eq:lagrange}; the
$\hat{\sigma}({\bf r})$ delta function constraint is expressed with
auxiliary fields $\psi^{ij}({\bf r})$, and the disorder average is
performed non-perturbatively with the replica trick
\cite{kn:gnBinder}; replica symmetry breaking (RSB) is not expected
and the replica limit is exact; still the disorder average yields 
a free energy functional non-local in polymer modes, and the GE approach
\cite{kn:gnEdwards} cannot be used.

The partition function in eq. \ref{eq:partition} may be also viewed as
the path integral for a quantum Green function in
imaginary-time-Lagrangian-representation. For computational
convenience, we exchange the Lagrangian to a Hamiltonian
representation \cite{kn:gnSwanson}, and the free energy is computed
variationally in the Hamiltonian reference frame:
\begin{eqnarray}
H_{\alpha} =\sum_{\alpha} (-\frac{1}{4 A} \hat{\bf p}^{2}_{\alpha} 
+h_{\alpha}\hat{\bf u}^{2}_{\alpha}) \nonumber\\ 
h_{\alpha} = \frac{3(1-\overline\theta)}{4l} +
\frac{(1+\overline\theta)(\lambda+\psi_{\alpha})}{2};
A=\frac{\beta \epsilon (1 + \overline{\theta})}{4}
\label{eq:Hamiltonian}
\end{eqnarray}
$\hat{\bf p}^{2}_{\alpha}$ and $\hat{\bf u}^{2}_{\alpha}$ are
the $\alpha$ quantum kinetic energy and coordinate  
operator in imaginary time with $\alpha=x,y,z$.
$\psi_{\alpha}$ is the principal axis representation of the fields
$\sqrt{-1} \psi^{ij}$ while $\lambda$ equals $\sqrt{-1}$ times the auxiliary field that sets the 
magnitude of ${\bf u}(n_{i})$ to 1 for the A segments in eq. \ref{eq:partition}.
Based on eq. \ref{eq:Hamiltonian}, creation/annihilation operators, ($a^{+},a$), 
are introduced to facilitate computation of orientational 
averages in the free energy.
\begin{eqnarray}
\hat{\bf u}_{\alpha}=\frac{a_{\alpha}+a^{+}_{\alpha}}{\sqrt{2m
\omega_{\alpha}}} \; ;\;\hat{\bf
p}_{\alpha}=(a_{\alpha}+a^{+}_{\alpha})\frac{\sqrt{m
\omega_{\alpha}}}{2}\nonumber\\ m=2A\; ; \;
\omega_{\alpha}=(\frac{h_{\alpha}}{A})^{\frac{1}{2}}:
[a_{\alpha},a^{+}_{\beta}]=\delta_{\alpha,\beta}
\end{eqnarray}
Omitting presentation of the intermediate calculation \cite{kn:gnjcp_paper}, 
the free energy per segment obtained is:
\begin{eqnarray}
F=\frac{1}{2}\sum_{\alpha}(\frac{h_{\alpha}}{A})^{\frac{1}{2}} +
\frac{\overline{\delta \theta^{2}} 3}{64 A^3}\sum_{\alpha}h_{\alpha} + \frac{\overline{\delta \theta^{2}}
C}{32 A^{2}}\sum_{\alpha} S_{\alpha} \nonumber\\ - \frac{\overline{\delta \theta^{2}}}{64
A}\sum_{\alpha}\frac{ S_{\alpha}^{2}}{h_{\alpha}}-\lambda k + 
\frac{1}{\rho}(\frac{w}{2}[(\sum_{\alpha} \sigma_{\alpha})^{2}\nonumber\\
- \sum_{\alpha}\sigma_{\alpha}^{2}] 
- \sum_{\alpha} \psi_{\alpha} \sigma_{\alpha})\nonumber\\\label{eq:free}
\end{eqnarray}
\begin{eqnarray}
S_{\alpha}=-\frac{3}{2l} + \psi_{\alpha} +\lambda\; ;\;
C=\frac{\beta \epsilon}{2}; k=\frac{(1+\overline{\theta})}{2}\nonumber\\
\label{eq:parfree}
\end{eqnarray}
$\rho$ is the segment density;
$\sigma_{\alpha}$ is $\sigma_{i,j}$ in principal axis representation:
\begin{eqnarray}
{\sigma}^{\alpha}= \left( \begin{array}{ccc} a-b & 0 & 0\\
0 & a+b & 0\\ 0 & 0 & 2a \end{array} \right)  \label{eq:Vintmatrix}
\end{eqnarray}
For uniaxial ordering the orientational order parameter, $<S>$, is
given by $<S> = \frac{-3a}{b}$.  $1>\hspace{0.4cm}<S>\hspace{0.4cm}>0$
signals uniaxial nematic ordering while
$-0.5<\hspace{0.4cm}<S>\hspace{0.4cm}<0$ signals discotic ordering.
In Einstein notation, in the isotropic phase, ${\sigma^{ii}
(\bf{r}})=\rho_{A}({\bf r})$, otherwise, ${\sigma^{ii}
(\bf{r}})\neq\rho_{A}({\bf r})$.  The homopolymer limit of
the free energy, i.e. f=1 in eq. \ref{eq:free}, reproduces exactly
the GE free energy of a many-chains LCPs (viz. eq. 32 in
\cite{kn:gnEdwards}), and the minimal value for uniaxial ordering,
$<S>$=0.25,  obtained analytically in \cite{kn:gnEdwards}), is reproduced herein. 
\begin{figure}[ht]
\vspace{-1.5cm}
\psfig{file=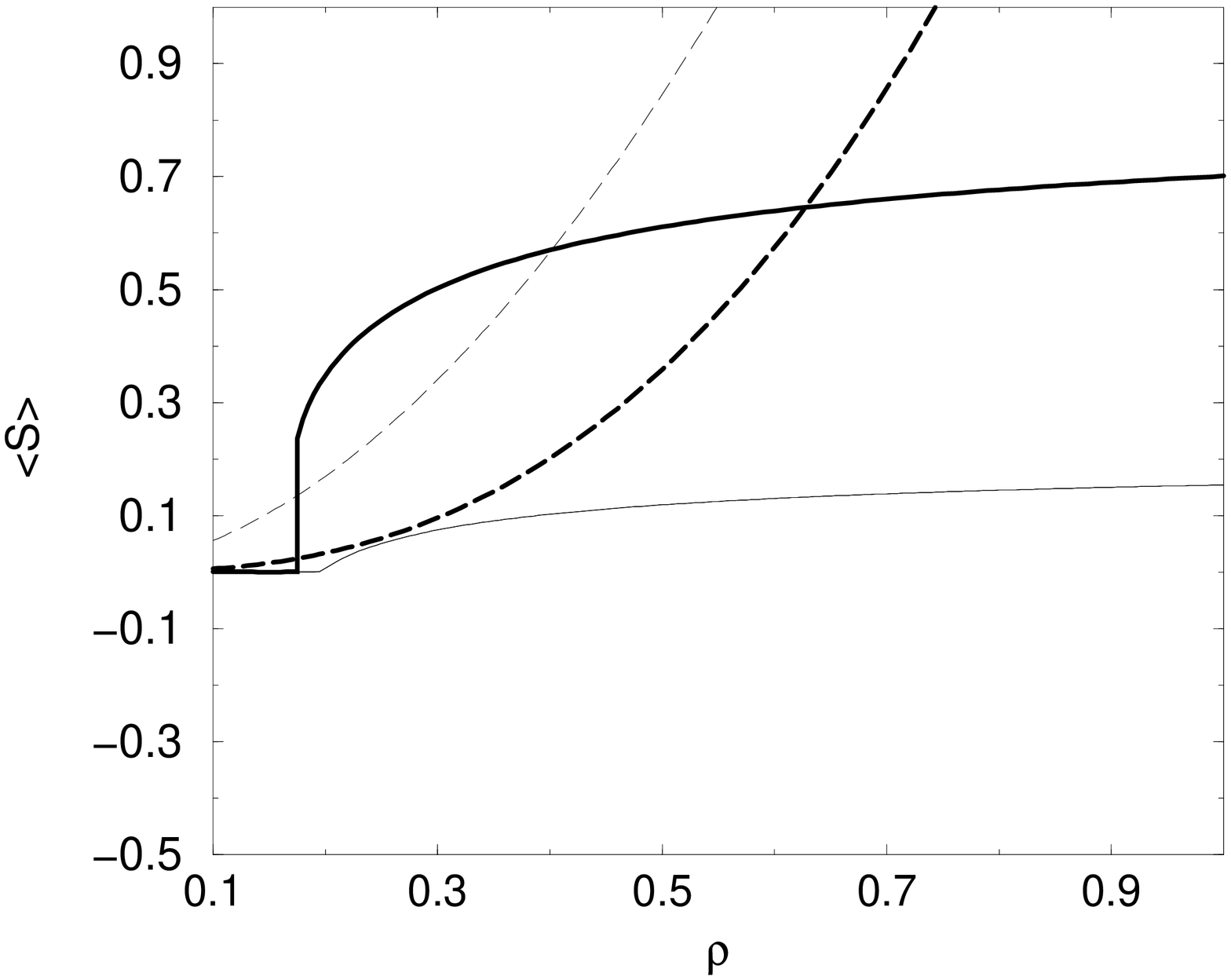,width=7.cm,angle=270}
\vglue-0.2cm
\vspace{0.4cm}
{\small Fig. 2: Numerical results for uniaxial ordering $<S>$ (solid lines) 
and chemical potential
per segment (long dashed lines) variation with segment density, $\rho$. 
Comparison among the thermodynamical stable phase (thick line) and the unstable
phase (thin line); f=0.938, w=10, $\beta e$=5, l=8.}
\label{cv71}
\end{figure}
\vglue.01cm
\vspace{-0.1cm}
Now, the free energy given in eq. \ref{eq:free} is minimized
analytically by the fields $\sigma_{\alpha}, \psi_{\alpha}, \lambda$
with $\alpha=x,y,z$;
seven non-linear self-consistent equations are obtained 
and solved numerically by direct iteration with proper mixing of
coefficients \cite{kn:gncarnahan}; more computational
and analytical details will be given elsewhere \cite{kn:gnjcp_paper}.
Fig. 2. depicts numerical results for lyotropic
stability of disordered LCP at small disorders. Thick lines in fig. 2
are uniaxial ordering curves for $<S>$ while dotted lines are the
chemical potentials of the stable phases.
\begin{figure}[ht]
\vspace{-1.5cm}
\psfig{file=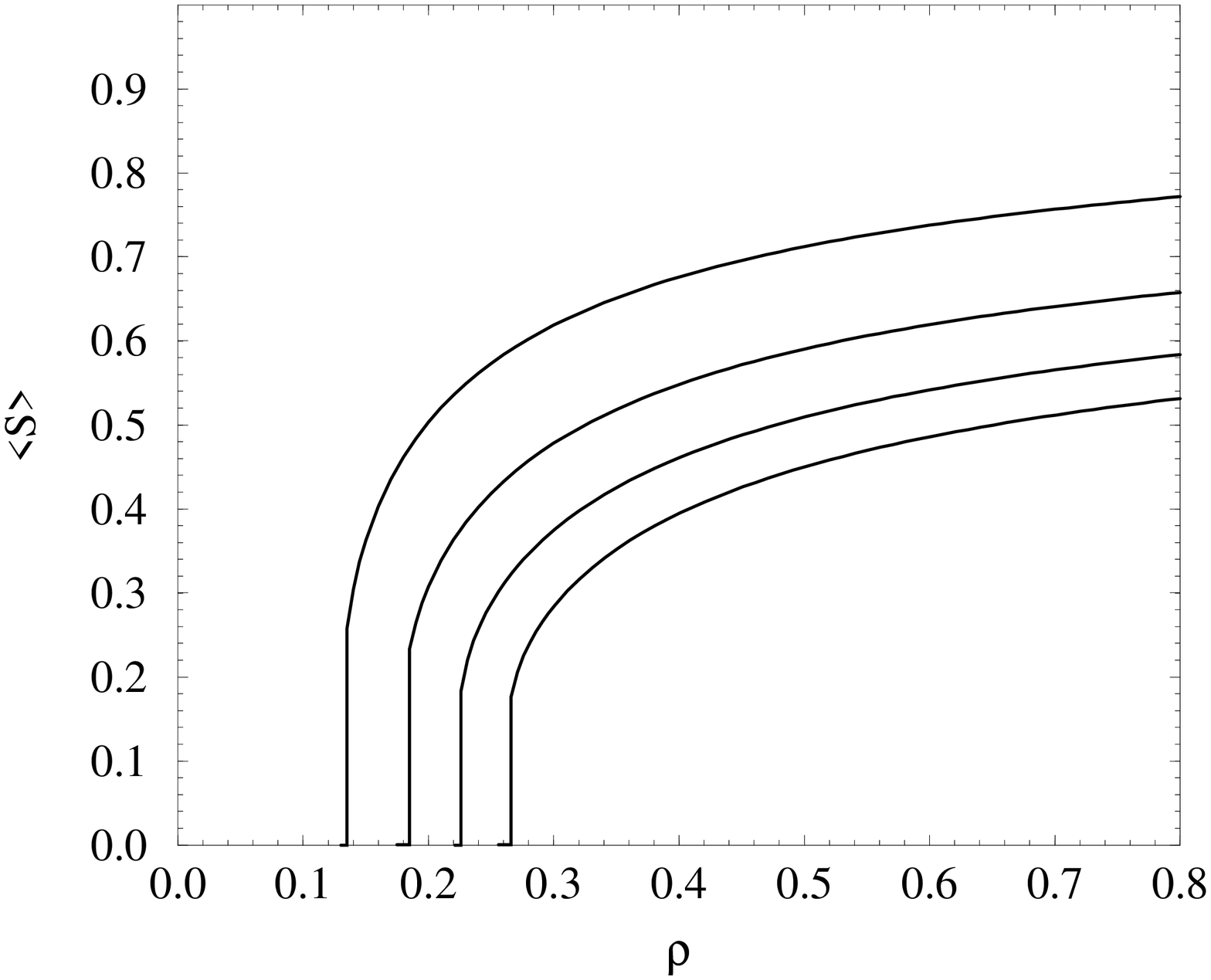,width=7.cm,angle=270}
\vglue-0.2cm
\vspace{0.4cm}
{\small Fig. 3: 
Numerical results for uniaxial ordering $<S>$ variation with segment density, $\rho$ and 
composition, f; w=10, $\beta e$=5, l=8; f values follow upper to lower $<S>$ curves;
f=0.98, 0.94, 0.93, 0.88.}
\label{cv72a}
\end{figure}
\vglue.01cm
\vspace{-0.1cm}
The thick line showing a discontinuous jump in $<S>$ at a finite density 
depicts a first order N/I transition; fig. 2 depicts also another stable solution
for $<S>$ (viz. other thick line); this solution increases continuously with increase in 
density from $<S>$=0, i.e. a second or higher order N/I transition.

The first order line bears a lower chemical potential
at all densities studied (viz. thick dashed line), and describes the physical ordering
scenario that should be observed in simulation or experiments. The
weak N/I transition, is also numerically stable,
but has a higher chemical potential (viz. thin dashed line) and it should not be observed.

Fig 3. is a numerical study of sequence heterogeneity effects on
lyotropic nematic ordering. Only stable solutions with lowest chemical
potentials are plotted.  Fig. 3 depicts $<S>$ - the nematic order
parameter variation with density for decreasing fractions of the stiff
segments, f.  The entropy carried by the flexible runs have dramatic
effects on ordering; Fig. 3 shows that a relatively small increase in
the average number of flexible segments decreases notably the overall
uniaxial ordering, and shifts the segment density at the N/I threshold
to higher values.  Interestingly, while the presence of sequence
heterogeneity impacts strongly the value of the
segment-density-N/I-threshold, the presence of disorder carried by the
fluid does not change the order of the N/I transition, and the
transition remains first order.
\begin{figure}[ht]
\vspace{-1.5cm}
\psfig{file=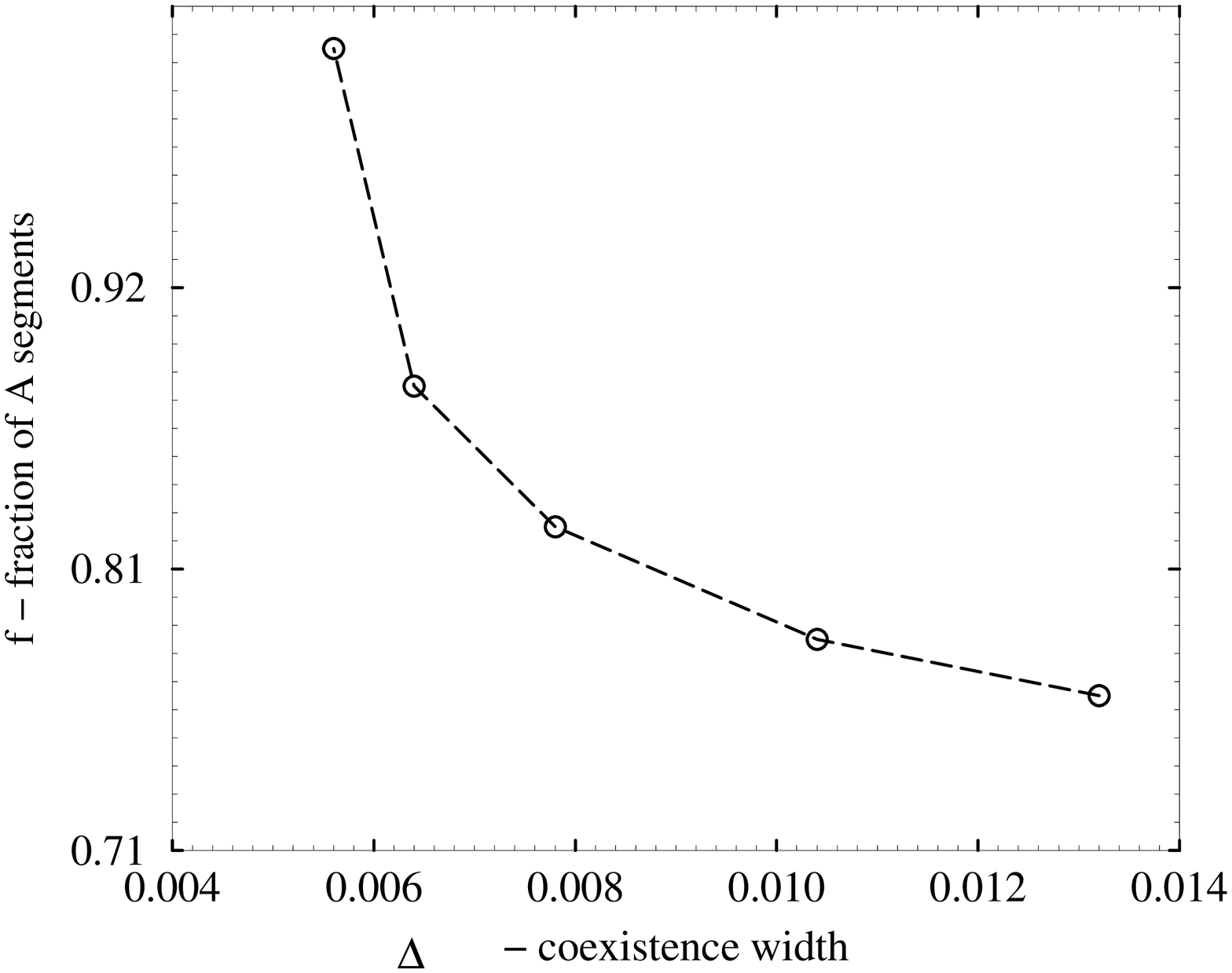,width=7.cm,angle=270}
\vglue-0.2cm
\vspace{0.4cm}
{\small Fig. 4: Numerical results for N/I biphasic width 
variation with fraction of mesogen segments, f; w=10, $\beta e$=5, l=8.}
\label{cv73a}
\end{figure}
\vglue-.02cm
\vspace{-0.4cm}

Recently Radzihovsky et. al. \cite{kn:gnRadzi} 
have shown that weak external disorder is sufficient to destabilize the 
nematic phase in LC rods; in the present system, wherein the disorder is carried by the 
fluid, nematic ordering is also affected; our calculation show that
above a critical value of disorder strength, the nematic ordering is destroyed 
and the phase becomes isotropic, yet the overall effect on uniaxial ordering
is less pronounced then the effect of external disorder. 

A Maxwell construction is employed (viz. \cite{kn:gnEdwards}) for the
calculation of the N/I coexistence region.  Free energies and chemical
potentials in the vicinity of the N/I transition are computed
numerically for the isotropic and the nematic phases for several
fractions of mesogenic segments.  $\Delta$, the N/I coexistence width
is given by $\Delta=\rho_{i}-\rho_{n}$; $\rho_{i}$ is the segment
density in the isotropic phase while $\rho_{n}$ is the segment density
in the nematic phase at coexistence.  Fig. 4 is a study of sequence
heterogeneity effects on the N/I coexistence.  In fig. 4 numerical
results for the coexistence width dependence on mesogenic fraction are
depicted.  Larger sequence heterogeneities increase the N/I
coexistence width almost 3 folds that of a semi-flexible homopolymer.

\begin{figure}[ht]
\vspace{-1.5cm}
\psfig{file=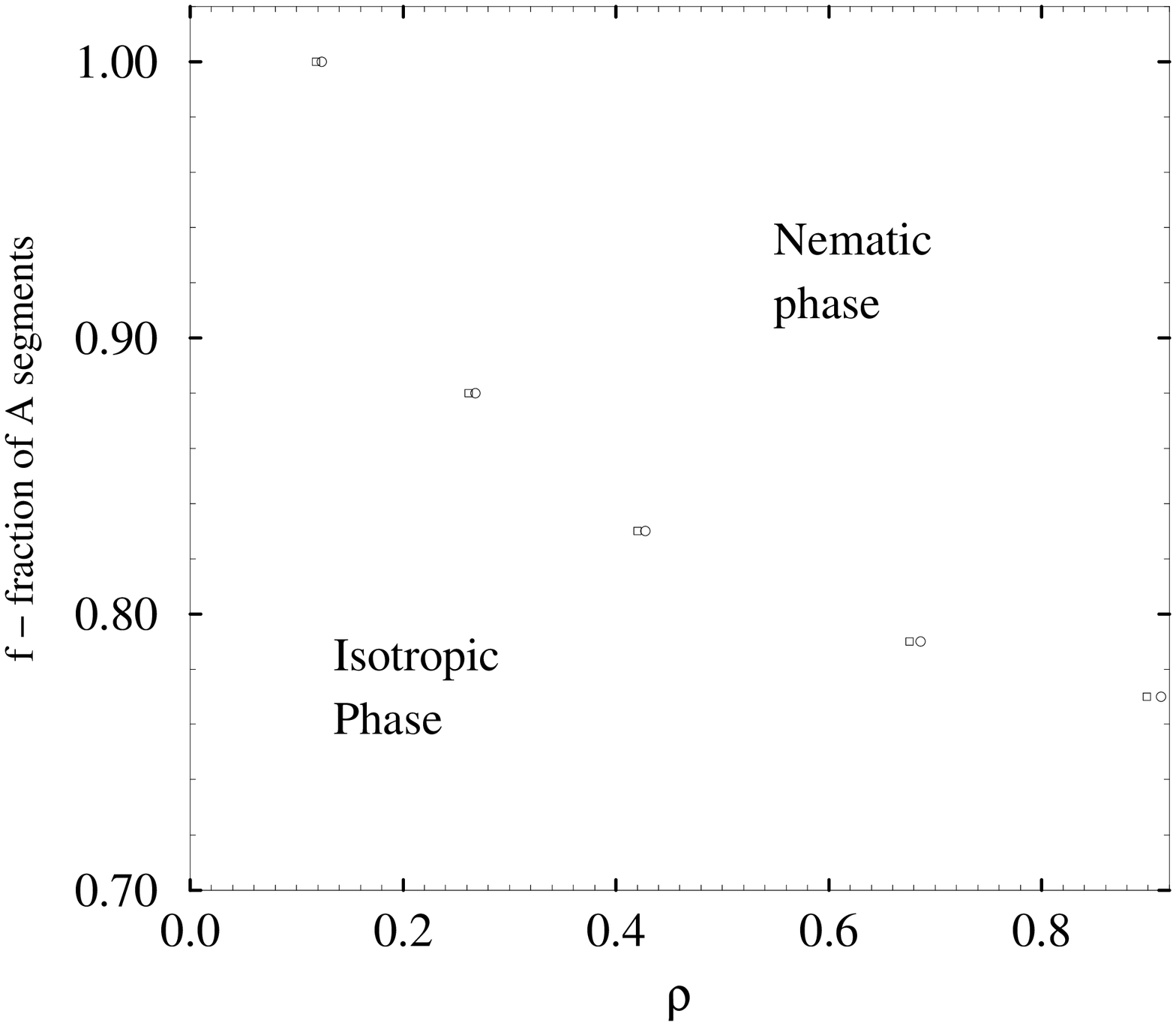,width=7.cm,angle=270}
\vglue-0.2cm
\vspace{0.4cm}
{\small Fig. 5: N/I phase diagram in f and $\rho$ variables;  w=10, $\beta e$=5, l=8;
 squares are (f,$\rho_{i}$) points, circles are (f,$\rho_{n}$) points.}
\label{cv3b}
\end{figure}
\vglue.01cm
\vspace{-0.4cm} Fig. 5 depicts the effect of sequence heterogeneity on 
orientational ordering. The entropy carried by mesophases
rich in flexible segments destroys the nematic ordering below a specific fraction
of mesogen segments (which is density dependent), the nematic phase loses stability, 
while the isotropic phase remains stable. For the values of parameters in fig. 5
at melt densities, ($\rho=1$), the critical fraction of mesogen segments, f, is $f_{c}$= 0.76. 
The N/I coexistence width displayed in fig. 5 as the horizontal distance among an
adjacent square ($\rho_{i}$) and circle ($\rho_{n}$) is  marginal at all densities displayed.

Invoking the homogeneous isotropic limit, (i.e. the difference in
anisotropy of A and B segments is negligible) the sequence
heterogeneity effects vanish in the infinite chain limit in agreement
with \cite{kn:gnFred}; In many experimental scenarios,
(i.e. $(MBPE-(methylene)_{y}$ with y=17, 18, 20 \cite{kn:gnpercec}) one
segment is stiff, the other is flexible and rigorous treatment of
anisotropy dissimilarity of A and B segments developed herein is crucial. In the
present work, we showed that for long chains, $\frac{\beta
e}{2}/{L}<<1$, the sequence heterogeneity effect on the N/I biphasic
region does not vanish, but in most cases is small with a $\Delta T
\sim 10^{o}$. contribution to the N/I biphasic width.  Our predictions
suggest that optical mesophases in the regime $\frac{\beta
e}{2}/{L}<<1$ cannot be associated with a wide N/I biphasic regime
induced by sequence-heterogeneity, and a different explanation of
experiments, based on a rigorous account of the optical scale ordering of
mesophases (e.g. $\sigma ({\bf k}) \neq 0$), is necessary.  The
occurrence of optical domains observed in experiments,
\cite{kn:gnStupp2}, can also be a signature of slow ordering dynamics
during a finite heating/cooling rate process. Currently, we are
studying both scenarios.

{\bf Acknowledgment} We gratefully acknowledge Harvard
University for financial support.

\end{multicols}
\end{document}